\documentclass{ws-p8-50x6-00}
\usepackage{epsfig,epsf,rotating}
\begin{document}

\title{SOFT PARTICLE PRODUCTION AT HERA}

\author{A.Rostovtsev}

\address{Institute f. Theoretical and Experimental Physics, ITEP, \\
B. Cheremushkinskaja 25, Moscow 117259, Russia\\   
E-mail: rostov@iris1.itep.ru}


\maketitle

\abstracts{
HERA data on soft hadron production are discussed. The
measured inclusive and exclusive hadron cross sections show a
simple scaling behaviour.
}

The final state hadrons produced in collision of high energy particles 
carry information about the underlying dynamics of the interaction.
Different theoretical approaches have been adopted to describe
such hadroproduction. Generally, the applicability of these theoretical
calculations depends on the type of produced hadrons and the value of
hadron
transverse momentum~($p_t$).
In practice, to describe the production of hadrons within whole
kinematic range, phenomenological models such as the Lund string 
model\cite{Lund}
are used. These models, however, 
contain a large number of parameters, values of which are not known {\it a
priori} and have to be determined from experiment.
Experimentally, measurements of hadroproduction are available 
 up to ISR 
energies and in $e^+e^-$ collisions at LEP.  
At higher collision energy only the 
HERA experiments provide the detailed information about hadrons of 
different flavours produced within a broad range of $p_t$. 
In this paper, the HERA data on inclusive and exclusive production of
hadrons in
$\gamma{p}$ interactions at average $W\equiv\sqrt{s_{\gamma{p}}}\approx
200~GeV$ and $Q^2\approx 0$ (so-called, photoproduction) are discussed. 

Recently, H1 has measured the inclusive $p(\bar{p})$ photoproduction
cross section. 
In H1 the protons are identified using
$dE/dx$ measurement in the central drift chamber. Thus, the proton data
are limited in momentum $(0.3<p_t<0.55~GeV)$ and rapidity
$(-0.3<y<0.3)$ range.
The measurements are made in the laboratory frame of reference and are
presented as the average cross-sections $E {d^3\sigma/d^3p}$ 
for the photoproduction
of protons and antiprotons, that is, half the sum of the proton and
anti-proton cross
sections. Here $E$ and $p$ are the proton energy and
momentum, respectively.
This cross-section in Fig.1 is shown as a function of the
particle transverse momentum  $p_t$. The errors shown are the
quadratic sum of the statistical and systematic uncertainties.
The systematic component is dominant. 
\begin{figure}[t]
\epsfxsize=18pc 
\begin{center}
\epsfbox{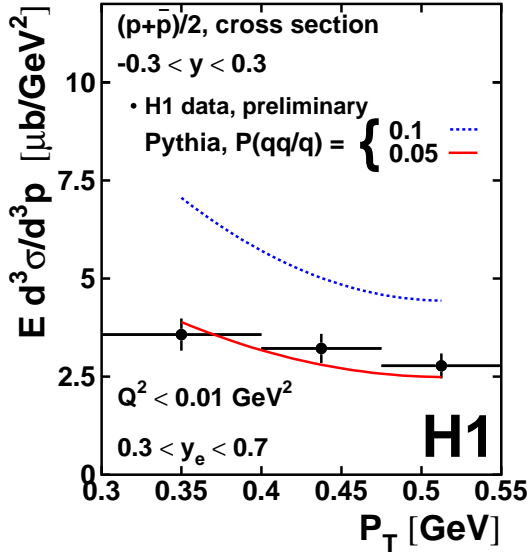} 
\caption{
The measured cross-section $E {d^3\sigma \over
d^3p}$ as a
function of transverse momentum, $p_t$, for rapidity values in
the range $-0.3<y<0.3$.
The cross-section represents half of the sum of protons and
anti-protons inclusively produced in
photoproduction interactions. Calculations of the PYTHIA  model are shown
with different values
of the diquark suppression factor of 0.1 (upper curve) and 0.05 (lower
curve).
  \label{fig:proton}}
\end{center}
\end{figure}

The production of baryons  within the Lund model is
regulated by several parameters, the most important of which in the
context of this measurement being the diquark suppression factor 
$P(qq/q)$ which
determines the probability of producing a $(qq, \bar{qq})$ pair
relative to a $(q,\bar{q})$ pair in the colour field. Results from
$e^+e^-$ collisions at LEP favour a value of $P(qq/q)$ approximately
0.1\cite{eestudy}.

The predictions of the Lund model are presented using diquark
suppression factors of 0.1 and 0.05. The model
over-estimates the data with the $e^+e^-$-derived value of 0.1 although
it provides a fair description the shape of the spectra. However,
a  description of both shape and yield is provided if a
diquark suppression value of $0.05$ is used.
This could signal an absence of the universality of baryon
production within this approach. However, more
studies are required of different baryons over a wider range of
phase space range than covered in this work in order to further 
investigate this.

The measurement of the proton inclusive cross section completes the
series of measurements on
the long-lived low mass hadrons $\pi$\cite{h1:pi}, $K$,
and $\Lambda$\cite{h1:K}. The pion data are recalculated from
the measured charged particle spectra by reducing these spectra by $17\%$
to take in to account an admixture of kaons and protons. 
In Fig.2 the low mass hadron production cross sections are plotted as
function of $p_t$. To make the comparison of different spices of hadrons
the cross sections are given for 
the particle's one isospin and one spin projection. This approach
follows the one used to compare the hadron yields in $e^+e^-$
collisions\cite{scaling2}.  
\begin{figure}[h]
\begin{center}
\hspace*{0.0cm}
\epsfig{
file=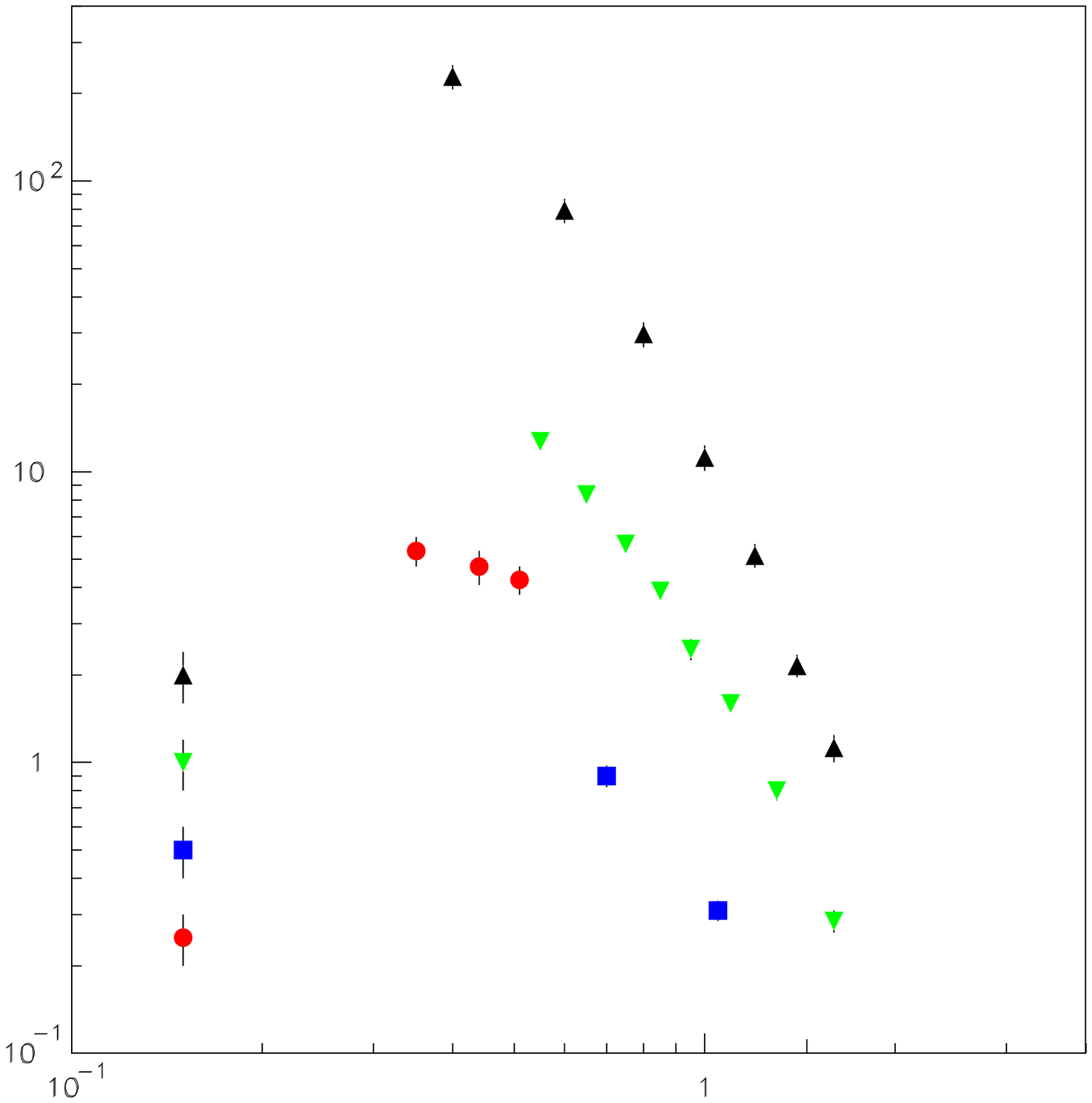,
        height=8.0cm,angle=0}
\put(-156,85){\large\bf $\pi^+$}
\put(-156,70){\large\bf $K^0$}
\put(-156,55){\large\bf $\Lambda$}
\put(-156,40){\large {\bf $p$} H1 prelim.}
\put(-50,180){\Large\bf H1}
\put(-80,165){\small\bf $\sqrt{s_{\gamma{p}}}\approx 200~GeV$}
\put(-80,0){\large\bf $P_t~[GeV]$}
\put(-220,35){\begin{sideways}\Large\bf
$\frac{1}{(2J+1)}\frac{d^2\sigma}{dydP_t^2}_{y_{lab}\approx 0}
[\mu{b}/GeV^2]$\end{sideways}}
\caption{
The measured inclusive double differential
hadron cross-section $d^2\sigma/dydP_t^2$
function of transverse momentum, $p_t$, for rapidity values $y=0$ as
measured in lab. system.
The cross-sections are given for 
the particle's one spin and isospin projection. 
  \label{fig:incl1}}
\end{center}
\end{figure}

As seen from the figure 2, the behaviour of the individual particle cross
sections at low values of $p_t$ 
seems to depend strongly on the specific hadron
mass. It was shown ({\it e.g.} V.Khoze~{\it et al}\cite{Khoze}), that in
QCD in the limit of $p\rightarrow 0$ the
inclusive hadron yield is defined by the hadron mass~$(m)$, which 
plays a role of an infrared cutoff in the QCD calculations.
Moreover, it was found at LEP, that the production rates of hadrons are
determined only by particle spins, isospins and their masses.
On the other hand, in high energy hadronic collisions the inclusive
particle spectra are described\cite{h1:pi,ua1} by a simple function 
$A(p_0+p_t)^{-n}$. It is tempting to assume
an intrinsic relation between $p_0$ and~$m$. 
We test this conjecture with the HERA data and 
replot the cross section data as function of $m+p_t$ (see Fig.3). 
\begin{figure}[h]
\begin{center}
\hspace*{0.0cm}
\epsfig{
file=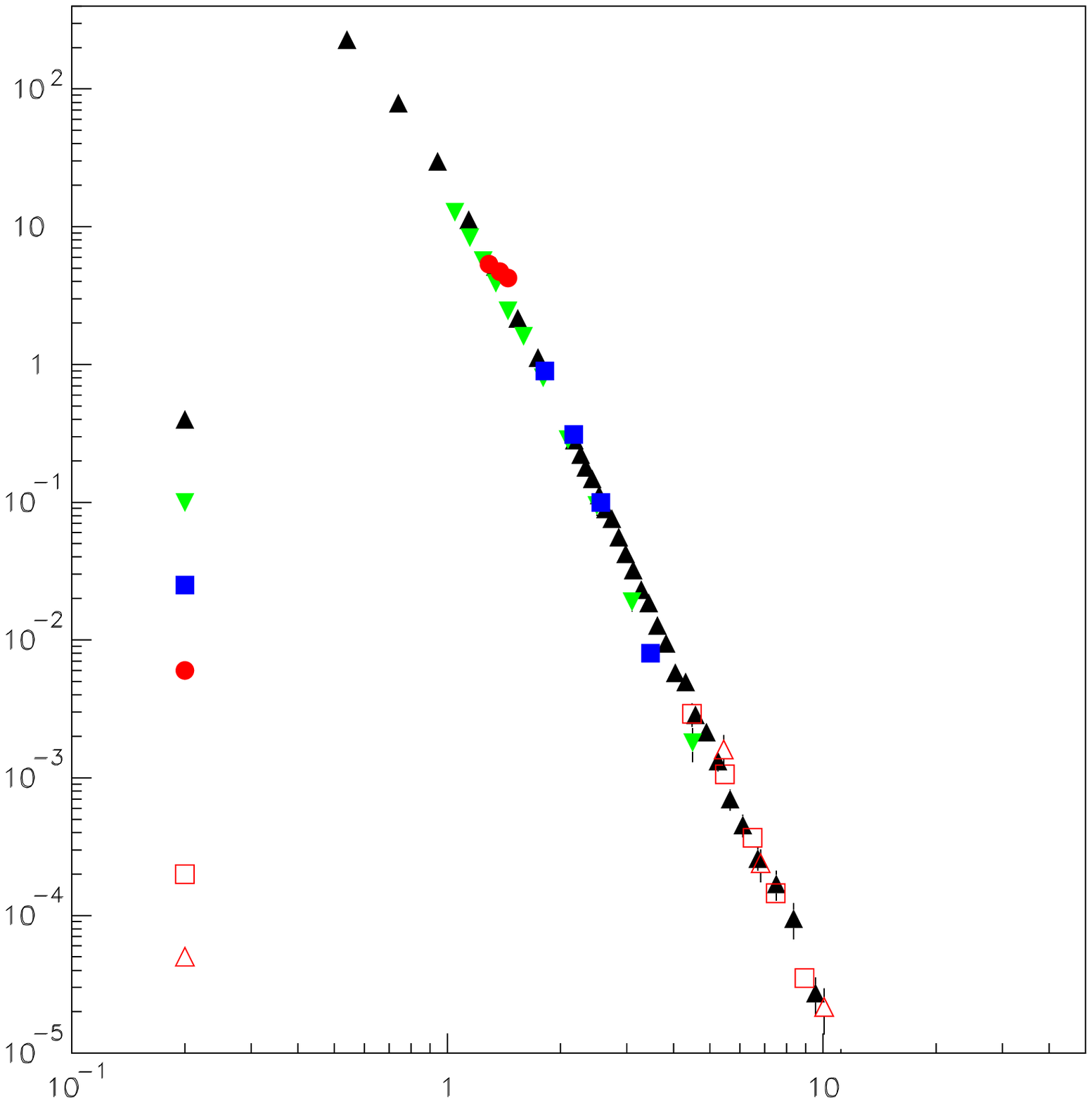,
        height=8.0cm,angle=0}
\put(-156,132){\large\bf $\pi^+$}
\put(-156,117){\large\bf $K^0$}
\put(-156,102){\large\bf $\Lambda$}
\put(-156,87){\large {\bf $p$} H1 prelim.}
\put(-156,50){\large\bf $D^{*+}$}
\put(-156,35){\large\bf $D_s^+$}
\put(-165,148){\Large\bf H1}
\put(-165,67){\Large\bf ZEUS}
\put(-75,185){\Large\bf HERA}
\put(-100,165){\large\bf $\sqrt{s_{\gamma{p}}}\approx 200~GeV$}
\put(-100,0){\large\bf $M+P_t~[GeV]$}
\put(-220,35){\begin{sideways}\Large\bf
$\frac{1}{(2J+1)}\frac{d^2\sigma}{dydP_t^2}_{y_{lab}\approx 0}
[\mu{b}/GeV^2]$\end{sideways}}
\caption{
The measured inclusive double differential
hadron cross-section $d^2\sigma/dydP_t^2$
function of transverse energy squared, $m+p_t$, for rapidity values
$y=0$ as measured in lab. system.
The cross-sections are given for
the particle's one spin and isospin projection. 
  \label{fig:incl2}}
\end{center}
\end{figure}
Surprisingly, the data shown in Fig.3
for different hadrons lay along one universal
curve. Adding the charm data\cite{charm} doesn't
spoil this picture. The inclusive $D^*$ and $D_s$ cross section data
impressively follow the  observed universal behaviour. 
The Tevatron data\cite{tev1} (not shown here) demonstrate
an approximate scaling for pions $K^0$ and $D^{*\pm}$ mesons when plotted
against the $m+p_t$. 

The H1 and ZEUS collaborations have also performed precision studies of 
the exclusive production
of vector mesons in the diffractive reaction 
\begin{eqnarray}
 \gamma + p \longrightarrow V + p \,\,\,\,(V=\rho^0, \omega,... \Upsilon)
\end{eqnarray}
Different assumptions and models prescriptions are used to describe the
meson yield from $\pi$ to $\Upsilon$. The low mass vector meson production
is traditionally described by Vector Dominance model, while $J/\Psi$
production is calculated in LO QCD with two gluon exchange. It was found,
that for heavier $\Upsilon$ one needs sizable corrections in addition to 
LO
QCD calculations. In addition, different mesons show different
dependences on $t$ and $W$. Given such a complicated theoretical cookery
the HERA data show an amazing scaling behaviour\cite{phi} similar to that
found for inclusive particle production. This universality is demonstrated
in Fig.4 where cross section for various vector mesons produced in the
reaction~1 at $W\approx 130~GeV$ and $Q^2\approx 0$ and
normalized by electronic meson width $\Gamma_{ee}$ are plotted
as function of $m^2$. The data demonstrate an approximately universal
$1/m^4$ behaviour. It is interesting to note the striking similarity 
with a scaling law
found by F.Halzen~{\it~et~al.}~1977 for narrow vector mesons $(\phi,
J/\Psi, \Upsilon)$ 
inclusively produced in $hh$ interactions\cite{Halzen}.
\begin{figure}[h]
\begin{center}
\hspace*{0.0cm}
\epsfig{
file=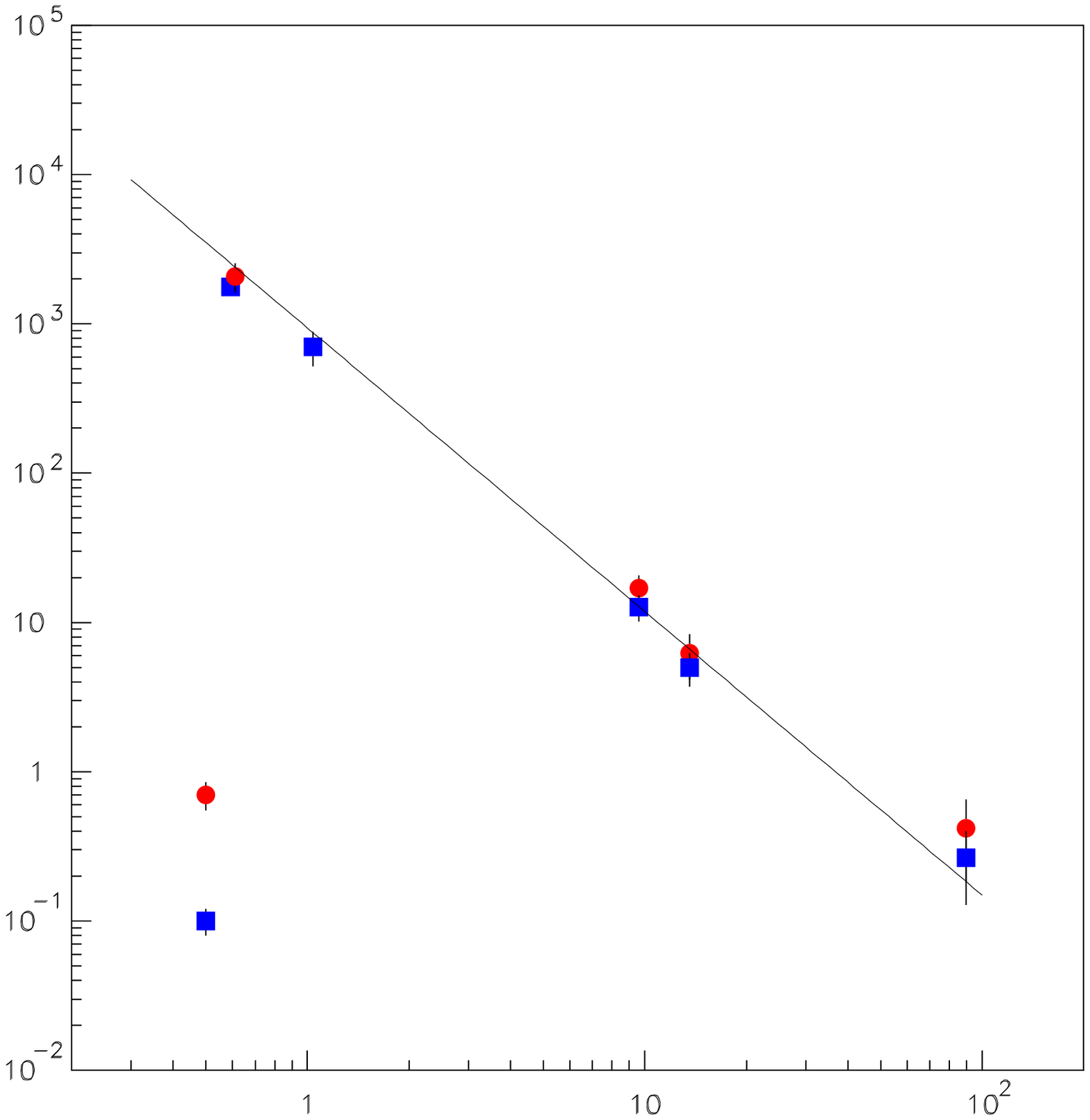,
        height=8.0cm,angle=0}
\put(-156,181){\large\bf $\rho^0$}
\put(-154,170){\large\bf $\omega$}
\put(-140,157){\large\bf $\phi$}
\put(-86,118){\large\bf $J/\Psi$}
\put(-73,101){\large\bf $\Psi(2S)$}
\put(-45,75){\large\bf $\Upsilon(1S)$}
\put(-142,98){\Large $\sim\frac{1}{M^4}$}
\put(-130,112){\bf \vector(1,1){15}}
\put(-145,68){\Large H1}
\put(-145,46){\Large ZEUS}
\put(-95,180){\normalsize\bf $\langle W \rangle =130~GeV$}
\put(-80,0){\large\bf $M^2~[GeV^2]$}
\put(-220,40){\begin{sideways}\Large\bf
$\frac{1}{\Gamma_{ee}}\sigma(\gamma{p}\rightarrow Vp)
[nb/keV]$\end{sideways}}
\caption{
The measured cross-section of exclusive vector meson photoproduction as
function of meson mass, $m$.
The cross-sections are normalized by meson electronic width $\Gamma_{ee}$. 
  \label{fig:excl1}}
\end{center}
\end{figure}

The data used in the Fig.4 include the recent measurements on high $W$
$\Psi(2S)$ (ZEUS) and $\omega$ (H1) production. The omega
cascade decay in the reaction~1 is reconstructed by detecting three
photons at very low angle w.r.t. to the electron beam axis. This
experimentally challenging task was accomplished by using the new low
angle calorimeter installed in H1. This devise was also used to search
the exclusive production of C=+1 mesons $(\pi^0, f_2$ and $a_2)$ decaying
into even number of photons. No signals for C=+1 states have been observed
and upper limits reported by H1 collaboration can be used to 
constrain the cross-section of Odderon-induced processes.

In summary, the HERA collider
experiments are rich and unique source of the data on
soft particle production. While different assumptions and models
prescriptions are used to describe the
meson yield from $\pi$ to $\Upsilon$ the data show an approximate scaling
behaviour 
Similar scaling was found in lower
energy data\cite{scaling1} and decay products of $Z^0$ at
LEP~\cite{scaling2}. 
Although the reason for this scaling is unknown this phenomenon is very
interesting and needs further studies. To this end, it is
important that HERA experiments provide further detailed information on
the hadronic resonance production.

\section*{Acknowledgments}
The work was partially supported by
Russian Foundation for Basic Research, grant 
RFBR-01-02-16431 and grant RFBR-00-15-96584.

\end{document}